\documentstyle[12pt,epsfig]{article}
\begin{document}
\def\bt{\begin{tabular}}
\def\et{\end{tabular}}
\def\bfr{\begin{flushright}}
\def\mm{\mbox{\boldmath $ }}
\def\efr{\end{flushright}}
\def\bfl{\begin{flushleft}}
\def\efl{\end{flushleft}}
\def\vs{\vspace}
\def\hs{\hspace}
\def\sta{\stackrel}
\def\pb{\parbox}
\def\bc{\begin{center}}
\def\ec{\end{center}}
\def\sp{\setlength{\parindent}{2\ccwd}}
\def\bp{\begin{picture}}
\def\ep{\end{picture}}
\def\uni{\unitlength=1mm}
\def\REF#1{\par\hangindent\parindent\indent\llap{#1\enspace}\ignorespaces}

\noindent \bc {\Large\bf Groundstate with Zero Eigenvalue\\
\vspace{.1cm} for Generalized Sombrero-shaped Potential\\
 \vspace{.3cm} in $N$-dimensional Space}

\vs*{1cm} {\large  Zhao Wei-Qin$^{1,~2}$}

\vs*{1cm}

{\small \it 1. China Center of Advanced Science and Technology
(CCAST)}

{\small \it (World Lab.), P.O. Box 8730, Beijing 100080, China}

{\small \it 2. Institute of High Energy Physics, Chinese Academy of
Sciences,

P. O. Box 918(4-1), Beijing 100039, China}

\ec
\vspace{2cm}
\begin{abstract}

Based on an iterative method for solving the goundstate of
Schroedinger equation, it is found that a kind of generalized
Sombrero-shaped potentials in N-dimensional space has groundstates
with zero eigenvalue. The restrictions on the parameters in the
potential are discussed.

\end{abstract}
\vspace{.5cm}

PACS{:~~11.10.Ef,~~03.65.Ge}\\


\newpage


{\large

Recently G. 't Hooft et al.[1] discussed the possibility of the
existence of a new kind of transformation from real to imaginary
space-time variables. The invariance of this transformation needs to
have zero eigenvalue for the groundstate. In this paper we discuss a
kind of generalized Sombrero-shaped potentials in $N$-dimensional
space and give restrictions to the parameters in the potential to
give zero eigenvalue for the groundstate. The generalized
Sombrero-shaped potentials in $N$-dimensional space are defined as
$$
V(r) = \frac{1}{2} g^2 (r^4-\alpha r^2+\beta)(r^2+A)\eqno(1)
$$
where $g^2$ and $\alpha$, $\beta$ and $A$ are taken to be
arbitrary constants. The corresponding Schroedinger equation for
the groundstate radial wave function is
$$
(-\frac{1}{2r^{N-1}}\frac{d}{dr}r^{N-1}\frac{d}{dr}+V(r))\psi(r)=E\psi(r)\eqno(2)
$$
The boundary conditions are
$$
\psi(\infty)=0~~~~~\psi'(0)=0.\eqno(3)
$$
In the following an iterative solution for the groundstate of (2) is
given.

For a groundstate it is reasonable to express the wave function as
$$
\psi(r)=e^{-S(r)}.\eqno(4)
$$
Substituting (4) into (2) an equation for $S(r)$ is obtained as
$$
S'(r)^2-\frac{N-1}{r}S'(r)-S''(r)=2(V(r)-E)\eqno(5)
$$
An iterative method has been developed by Friedberg, Lee and
Zhao[2] to solve this kind of problems. Following this iterative
method, we introduce a trial function
$$
\phi(r)=e^{-S_0(r)}\eqno(6)
$$
satisfying another Schroedinger equation
$$
(-\frac{1}{2r^{N-1}}\frac{d}{dr}r^{N-1}\frac{d}{dr}+V(r)-h(r))\phi(r)
=E_0\phi(r)=(E-\Delta)\phi(r),\eqno(7)
$$
where $h(r)$ and $\Delta$ are the corrections of the potential and
the eigenvalue of the groundstate. To ensure the convergency of the
iterative method it is necessary to construct the trial function in
such way that the perturbed potential $h(r)$ is positive (or
negative) and finite everywhere. Specially, $h(r)\rightarrow 0$ when
$r\rightarrow \infty$. Now substituting (6) into (7) we obtain the
equation for $S_0(r)$:
$$
S_0'(r)^2-\frac{N-1}{r}S_0'(r)-S_0''(r)=2(V(r)-h(r)-E_0).\eqno(8)
$$
Therefore
$$
h(r)+E_0=V(r)-\frac{1}{2}(S_0'(r)^2-\frac{N-1}{r}S_0'(r)-S_0''(r)).\eqno(9)
$$
For a finite $h(r)$ it should not include terms with positive power
of $r$. Since the highest order of $r$-power in the potential is $6$
and the potential $V(r)$ has only even powers of $r$, for arbitrary
normalization we can assume
$$
S_0(r)=(ar^4-cr^2)+m\log(r^2+1).\eqno(10)
$$
Substituting (10) into (8) to cancel $r^6$ term we have $a=g/4$ and
$$
S_0(r)=(\frac{g}{4}r^4-cr^2)+m\log(r^2+1).\eqno(11)
$$
To cancel the terms with $r^4$ and $r^2$ in (8) we get
$$
4c=(\alpha-A)g,\eqno(12)
$$
$$
4m=g(\beta- \alpha A)-\frac{1}{4}g(\alpha-A)^2+N+2.\eqno(13)
$$
and finally obtain
$$
h(r)=2m(m+1)\frac{1}{(r^2+1)^2}+((N-2)m-2m^2-2mg-4mc)\frac{1}{r^2+1}\eqno(14)
$$
$$
E_0=\frac{1}{2}Ag^2\beta+2mg-Nc+4mc.\eqno(15)
$$
When we set $m=0$, i.e. $h=0$, the trial function becomes the
exact groundstate wave function:
$$
\phi(r)=e^{-\frac{g}{4}r^4+cr^2}\eqno(16)
$$
with eigenvalue
$$
E_0=\frac{1}{2}Ag^2\beta-Nc.\eqno(17)
$$
This gives the following relation between the parameters:
$$
g(\beta- \alpha A)-\frac{1}{4}g(\alpha-A)^2+N+2=0.\eqno(18)
$$
Further introducing $E_0=0$, i.e.
$$
\frac{1}{2}Ag^2\beta-Nc=0\eqno(19)
$$
(12), (18) and (19) give the condition to the parameters for a
groundstate with zero eigenvalue.

When taking $c=0$, i.e. $A=\alpha$, we have
$$
E_0=\frac{1}{2}g^2A\beta\eqno(20)
$$
and there are two possible ways to give zero eigenvalue. For $A=0$
we have
$$
V(r)=\frac{1}{2}g^2(r^4+\beta)r^2\eqno(21)
$$
with its only minimum at $r=0$. For $\beta=0$, we have
$$
V(r)=\frac{1}{2}g^2r^2(r^4-A^2)\eqno(22)
$$
with its minima at $r=0$ and $r^2=A$. However, in both cases the
groundstate wave function is
$$
\phi(r)=e^{-\frac{1}{4}gr^4}\eqno(23)
$$
with its only maximum at $r=0$. Only when we take $c>0$, i.e.
$A<\alpha$, the wave function can have more than one maxima at
$r^2=2c/g$. For $c>0$ to ensure $E_0=0$ we must have
$$
g^2A\beta=2Nc>0.\eqno(24)
$$
Introducing $A=\eta \alpha$ and $\alpha^2=\lambda 4\beta$ the
condition of $m=0$ and $E_0=0$ gives the following two equations:
$$
g\beta(1-\lambda(1+\eta)^2)+N+2=0\eqno(25)
$$
$$
g\beta=\frac{1}{2}N\frac{1-\eta}{\eta}\eqno(26)
$$
which give a relation between $\eta$ and $\lambda$:
$$
\lambda N\eta^3+\lambda N\eta^2+(N+4-\lambda
N)\eta+N(1-\lambda)=0.\eqno(27)
$$
As an example taking $N=3$ the function $\eta(\lambda)$ is plotted
in Fig.~1 for $\lambda>1$ and $0<\eta<1$. For any chosen $g$ and
$\lambda$ the other parameters $\eta$, $\beta$, $\alpha$ and $A$ can
all be fixed to ensure $m=0$ and $E_0=0$. For the 3-dimensional
case, if choosing $g=1.5$ and $\lambda=1.5$, from (26) and (27) the
parameters are obtained as $\eta=1/3$, $\beta=2$, $\alpha=\sqrt{12}$
and $A=\sqrt{12}/3$. This gives $c=\sqrt{3}/2$. The groundstate wave
function is
$$
\phi(r)=e^{-\frac{3}{8}r^4+\frac{\sqrt{3}}{2}r^2}\eqno(28)
$$
which has maxima at $r^2=2c/g=2/\sqrt{3}$ and a valley at $r=0$. In
Figs.~2 and 3 the potential and the wave function are plotted
respectively. It is interesting to obtain these analytic solutions
of the groundstate wave function, which are degenerate and with zero
eigenvalue, by choosing special sets of parameters for the
potential.

As an example discussed by R. Jackiw one can introduce a parameter
$$
r_0^4=\frac{N+2}{3}.\eqno(29)
$$
When $g=1$, $\beta=r_0^4$ and $A=2r_0^2$ one can choose special
values of $\alpha$, which give $m=0$, to obtain analytic solutions
of the goundstate directly from the trial functions. In this set of
parameters the condition of $m=0$ gives the following equation:
$$
\alpha^2+4r_0^2\alpha-12r_0^4=0\eqno(30)
$$
which gives two solutions for $\alpha$: When taking
$\alpha=2r_0^2$ we have $c=0$ and the groundstate is
$$
\psi(r)=e^{-r^4/4},~~E_0=r_0^6,\eqno(31)
$$
which is the solution given by R. Jackiw in [3]; while
$\alpha=-6r_0^2$ gives $c=-2r_0^2$ and the solution is
$$
\psi(r)=e^{-r^4/4-2r_0^2r^2},~~E_0=r_0^6+2Nr_0^2.\eqno(32)
$$
If we write the potential in the form of
$$
V(r)=\frac{1}{2}g^2[(r^2-r_0^2)^2-\mu r_0^4](r^2-2\eta
r_0^2),\eqno(33)
$$
comparing (33) with (1) the parameters are related in the
following way:
$$
\alpha=2r_0^2,~~~\beta=r_0^4(1-\mu),~~~A=\eta \alpha.\eqno(34)
$$
To obtain an analytical solution for the groundstate with zero
eigenvalue, from (12), (18) and (19) we have
$$
c=\frac{1}{2}(1-\eta)gr_0^2,\eqno(35)
$$
$$
g(1-\mu)-g(1+\eta)^2+3=0\eqno(36)
$$
and
$$
\eta g^2(1-\mu)r_0^6-Nc=0.\eqno(37)
$$
The wave function is expressed as
$$
\phi(r)=e^{-\frac{g}{4}r^4+cr^2}\eqno(38)
$$
For $N=3$, taking $g=1$, substituting (35) and (36) into (37) an
equation for $\eta$ is obtained as following:
$$
2\eta^3+4\eta^2-\frac{11}{5}\eta-\frac{9}{5}=0.\eqno(39)
$$
From (39) and (36) we find values of $\eta$ and $\mu$, which give
$m=0$ and $E_0=0$, i.e. the analytical groundstate with zero
eigenvalue:
$$
\eta=0.797005,~~~~\mu=0.770765,
$$
correspondingly, from (35)
$$
c=0.103152~.
$$
The wave function has degenerate groundstate with zero eigenvalue
and its maxima at $r=\sqrt{2c/g}=0.4542$.\\

\noindent {\bf Acknowledgement }\\

The author would like to thank Professor R. Jackiw for introducing
the work by Professor G. 't Hooft. The author would also like to
thank Professor T. D. Lee for his continuous guidance and
instruction.

\vspace{1cm}

\noindent {\bf References}\\

1. G. 't Hooft and S. Nobbenhuis, Class. Quant. Grav. 23(2006)3819\\

2. R. Friedberg, T. D. Lee and W. Q. Zhao, Ann. Phys.
321(2006)1981\\

3. R. Jackiw, Private communication\\

\vspace{1cm}

\begin{figure}[h]
 \centerline{
\epsfig{file=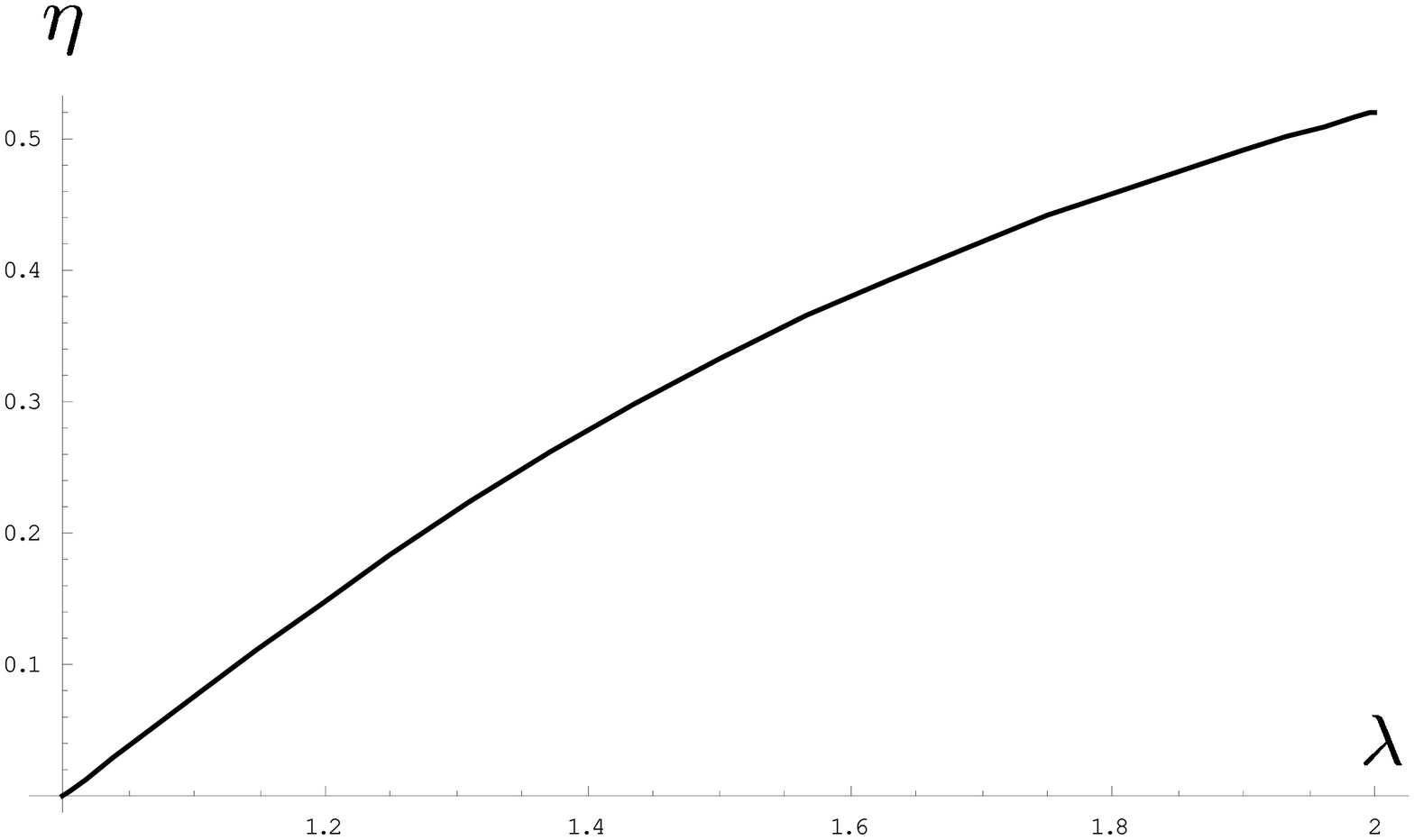, width=8cm, height=14cm, angle=-90}}
\vspace{.5cm}
 \centerline{{\normalsize \sf Fig. 1~  $\eta(\lambda)$ for the groundstate with zero eigenvalue }}
\end{figure}

\begin{figure}[h]
 \centerline{
\epsfig{file=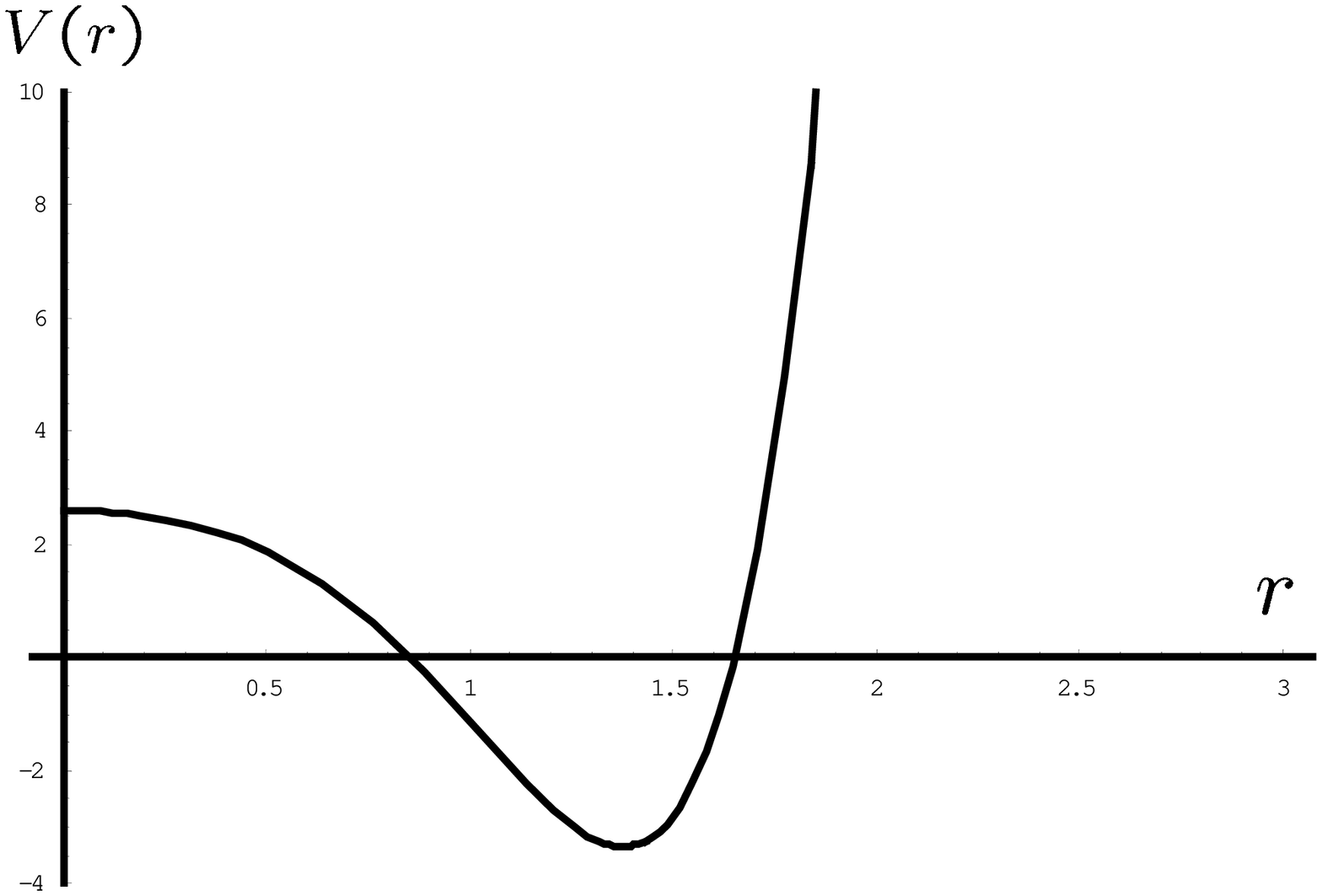, width=8cm, height=14cm, angle=-90}}
\vspace{.5cm}
 \centerline{{\normalsize \sf Fig. 2~ $V(r)$ for $N=3$, $g=1.5$ and $\lambda=1.5$}}
\end{figure}

\begin{figure}[h]
 \centerline{
\epsfig{file=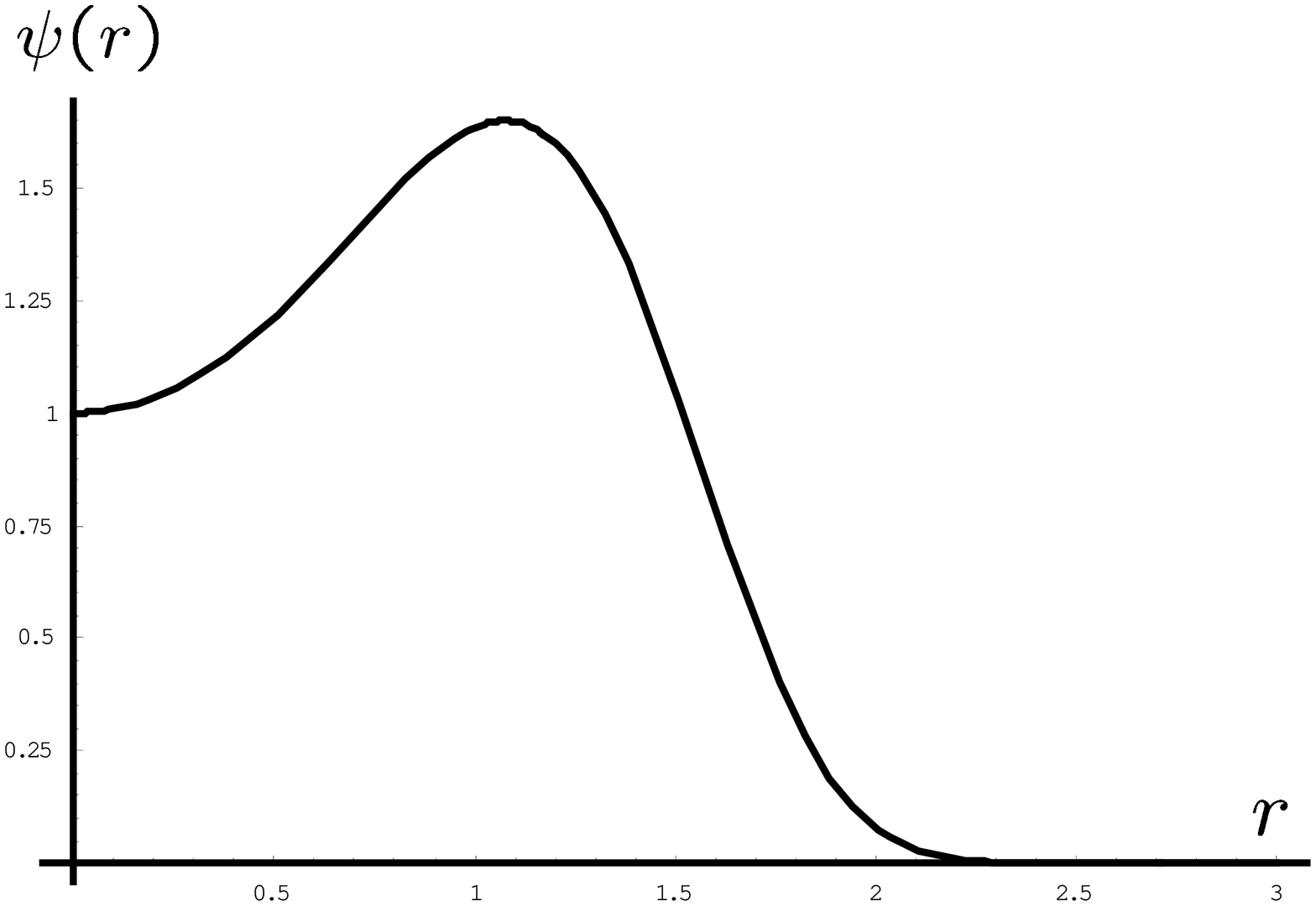, width=8cm, height=14cm, angle=-90}}
\vspace{.5cm}
 \centerline{{\normalsize \sf Fig. 3~ $\psi(r)$ for $N=3$, $g=1.5$ and $\lambda=1.5$ }}
\end{figure}

 }
\end{document}